\documentclass[sigconf]{acmart}

\AtBeginDocument{%
  \providecommand\BibTeX{{%
    \normalfont B\kern-0.5em{\scshape i\kern-0.25em b}\kern-0.8em\TeX}}}

\setcopyright{acmcopyright}

\copyrightyear{2023}
\acmYear{2023}
\setcopyright{acmlicensed}\acmConference[ETRA '23]{2023 Symposium on Eye Tracking Research and Applications}{May 30-June 2, 2023}{Tubingen, Germany}
\acmBooktitle{2023 Symposium on Eye Tracking Research and Applications (ETRA '23), May 30-June 2, 2023, Tubingen, Germany}
\acmPrice{15.00}
\acmDOI{10.1145/3588015.3589844}
\acmISBN{979-8-4007-0150-4/23/05}

\usepackage{subfigure}
\usepackage{enumitem}
\usepackage{hyperref}
\usepackage{tabularx}

\begin{document}

\title{Gazealytics: A Unified and Flexible Visual Toolkit for Exploratory and Comparative Gaze Analysis}
\author{Kun-Ting Chen}
\orcid{0000-0002-3217-5724}
\affiliation{%
  \institution{University of Adelaide}
  \country{Australia}
}
\affiliation{%
  \institution{University of Stuttgart}
  \country{Germany}
}
\email{Kun-Ting.Chen@adelaide.edu.au}

\author{Arnaud Prouzeau}
\orcid{0000-0003-3800-5870}
\affiliation{%
  \institution{Inria \& LaBRI (University of
Bordeaux, CNRS, Bordeaux-INP)
}
  \country{France}
}
\email{Arnaud.Prouzeau@inria.fr}

\author{Joshua Langmead}
\orcid{0009-0001-7801-992X}
\affiliation{%
  \institution{Monash University}
  \country{Australia}
}
\email{jlan0025@student.monash.edu}

\author{Ryan T Whitelock-Jones}
\orcid{0000-0002-5954-5692}
\affiliation{%
  \institution{Monash University}
  \country{Australia}
}
\email{ryanwhitelockjones@gmail.com}

\author{Lee Lawrence}
\orcid{0000-0002-1336-9136}
\affiliation{%
  \institution{Monash University}
  \country{Australia}
}
\email{Lee.Lawrence@monash.edu}

\author{Tim Dwyer}
\orcid{0000-0002-9076-9571}
\affiliation{%
  \institution{Monash University}
  \country{Australia}
}
\email{Tim.Dwyer@monash.edu}

\author{Christophe Hurter}
\orcid{0000-0003-4318-6717}
\affiliation{%
  \institution{ENAC, Université de Toulouse}
  \country{France}
}
\email{Christophe.Hurter@enac.fr}

\author{Daniel Weiskopf}
\orcid{0000-0003-1174-1026}
\affiliation{%
  \institution{University of Stuttgart}
  \country{Germany}
}
\email{Daniel.Weiskopf@visus.uni-stuttgart.de}

\author{Sarah Goodwin}
\orcid{0000-0001-8894-8282}
\affiliation{%
  \institution{Monash University}
  \country{Australia}
}
\email{Sarah.Goodwin@monash.edu}

\renewcommand{\shortauthors}{Chen and Prouzeau, et al.}
\newcommand{\webveta}{\textsc{Gazealytics}}

\newcommand{\rev}[1]{\textcolor{black}{#1}}
\newcommand{\kt}[1]{\textcolor[RGB]{255, 165, 0}{kt: #1}}
\newcommand{\slFOR}{{\bf{FOR}\normalfont}}
\newcommand{\slEVI}{{\bf{EVI}\normalfont}}
\newcommand{\slSCH}{{\bf{SCH}\normalfont}}
\newcommand{\slSEN}{{\bf{SEN}\normalfont}}
\begin{abstract}
We present a novel, web-based visual eye-tracking analytics tool called \webveta{}. 
\rev{Our open-source toolkit features a unified combination of gaze analytics features that support flexible exploratory analysis, along with annotation of areas of interest (AOI) and filter options based on multiple criteria to visually analyse eye tracking data across time and space}. 
\webveta{} features coordinated views unifying spatiotemporal exploration of fixations and scanpaths for various analytical tasks. 
A novel matrix representation allows analysis of relationships between such spatial or temporal features.
Data can be grouped across samples, user-defined AOIs or time windows of interest (TWIs) to support aggregate or filtered analysis of gaze activity.  
\rev{This approach exceeds the capabilities of existing systems by supporting flexible comparison between and within subjects, hypothesis generation, data analysis and communication of insights}.  
We demonstrate in a walkthrough that Gazealytics supports multiple types of eye tracking datasets and analytical tasks.



\end{abstract}




\maketitle


\section{Introduction}
\begin{figure*}
    \centering         \includegraphics[width=1.0\textwidth]{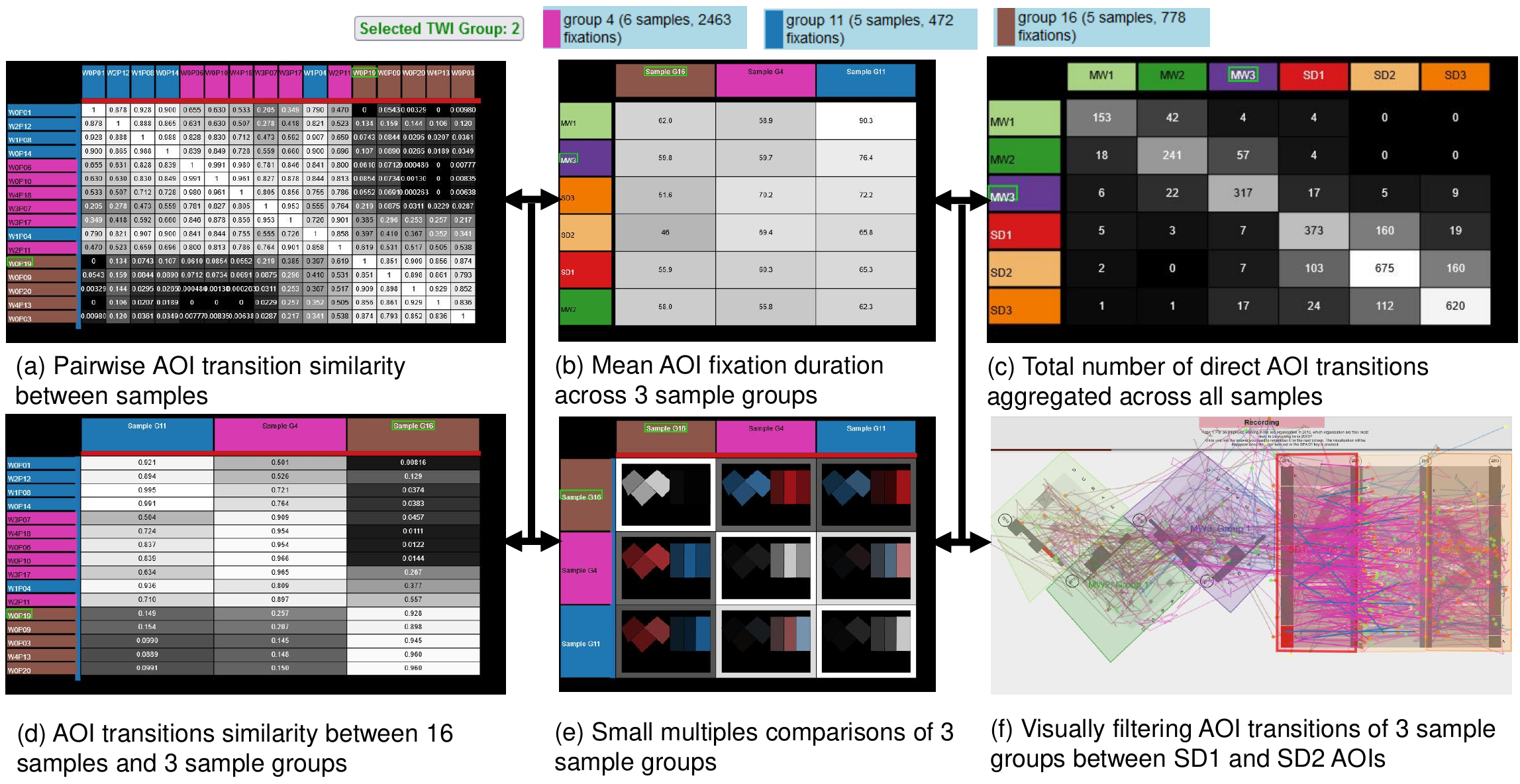}    
    \Description[]{}
    \caption[]{An example of multi-way visual exploration of a controlled experiment. \rev{An exploration can begin at any stage of (a--f) and move between them as shown by arrows (arrows between the leftmost and the rightmost columns are ignored for clarity). 
    Colours indicate 6 AOIs and the relationships between AOIs and sample groups (b). Sample groups were determined by their pairwise similarity comparison based on the number of AOI transitions (a).} 
    \label{fig:teaser}
    \vspace{-1em}
    }
\end{figure*}

Analysis of eye tracking data collected in observational studies is a complex process that aims to make sense of large amounts of gaze data involving  high-frequency gaze samples, collected from a number of individuals~\cite{pozdniakov2023teachers,afzal2022investigating}. Even in controlled experiments, there may be many samples collected, each involving multiple activities, which can result in long recordings requiring complicated multivariate data analysis~\cite{chang2018evaluation,servais2022attentional,burch2022eye}. 
Therefore, to make analysis tractable, researchers often identify key areas of interest (AOIs) within the field of view, as well as time windows of interest (TWIs) throughout the period of participant activity~\cite{menges2020visualization,kwok2019gaze}.


AOIs are typically pre-identified in the experimental setup and statistical methods can be used to test hypotheses concerning expected activity within and across AOIs.  
However, the comprehensive space-time data recorded from gaze tracking is not only useful for hypothesis testing but may also be used for building hypothesis, i.e., supported by data-driven exploratory eye tracking analysis~\cite{kurzhals2016visual}.

Many of the current tools are inflexible in the degree to which they support exploratory data analysis~\cite{menges2020visualization,burch2019interactive}. Existing non-visual tools allow analysis with a pre-defined hypothesis, but there is little in the way of tools that allow for the exploration or generation of hypothesis~\cite{Blascheck2016,panetta2020iseecolor}.
To our knowledge, there is no single tool that provides a highly dynamic visual exploration with filtering options based on multiple criteria over fixations, saccades and scanpaths across many ways of grouping the data, e.g., by samples, AOIs (spatial filtering), or TWIs (temporal filtering).  


\begin{figure*}
    \centering
    \includegraphics[width=0.9\textwidth]{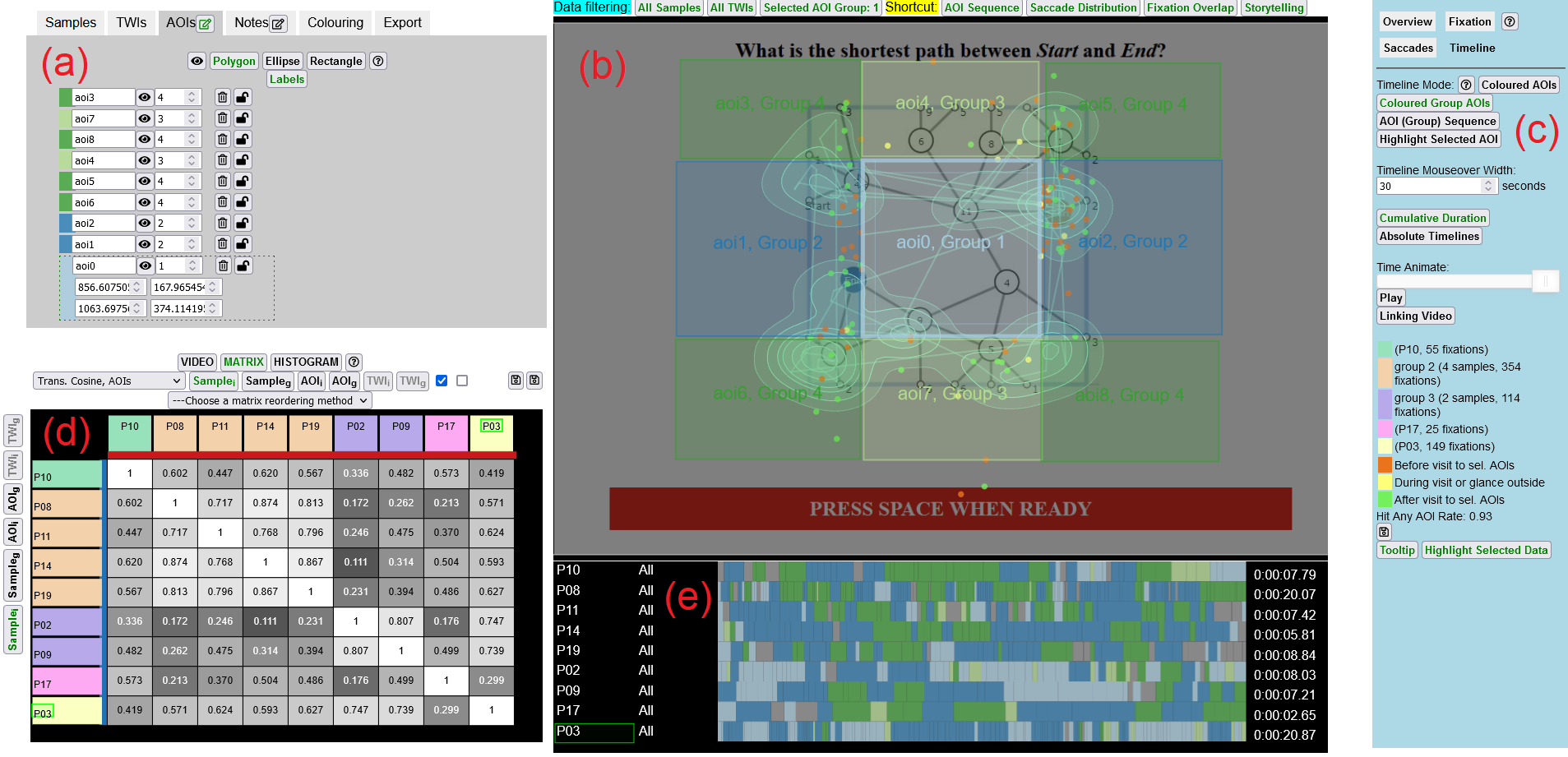}
    \Description[]{}
    \caption[]{Coordinated views of \webveta{}: (a) data management panel; (b) spatial panel; (c) parameter control panel; (d) metric panel; (e) timeline panel.
    }
    \label{fig:webveta_overview}
     \vspace{-1em}
\end{figure*}

Our contribution is twofold. First, we provide a multi-way visual exploratory approach for data from eye tracking studies, extensively using matrix-based overview (see~\autoref{fig:teaser}). As we summarise in~\autoref{sec:background}, a unified general matrix relationship approach has not been previously presented for eye tracking analysis.  
Our second contribution is an interactive group-level exploration tool for gaze data analysis. \webveta{} provides flexible visual data aggregation for identifying gaze patterns at various data dimensions: samples, AOIs, and TWIs. 
We show with eye tracking examples that \webveta{} supports flexible exploratory and comparative analysis.

We make \webveta{} publicly available, as an open-source gaze analytics toolkit\footnote{https://github.com/gazealytics/gazealytics-master}. 
\rev{The toolkit provides an easy-to-use interface to integrate into analysts' existing workflow}.
Several users have already used the web application for their real-world eye tracking data analysis problems (see, e.g., ~\cite{cai2022towards,pozdniakov2023teachers,chen2023reading}).

\section{Related Work}
\label{sec:background}

\begin{table*}
    \centering
    \caption{
    \rev{An overview of \webveta{} and existing analytical tools' support for eye tracking visualisation taxonomy (visualisation techniques) and 11 interactive visualisation task categories (grouped into Encode, Manipulate, Introduce). 
    The comparison criteria (3rd column) centres on metric visualisations and interactive visual grouping and exploration as a focus of this work.}
    }
    \vspace{-1em}    \includegraphics[width=1.0\textwidth]{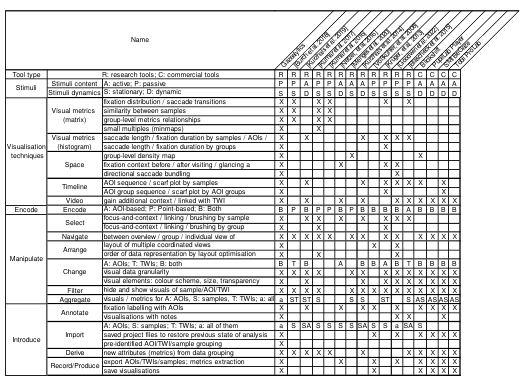}
    
    \label{tab:toolscompare}
     \vspace{-1em}
\end{table*}

\citet{Blascheck2017} survey techniques for interactive visualisation and visual analytics for eye tracking. Their taxonomy covers various facets of gaze analysis. In particular, they distinguish between static and dynamic stimuli, between passive and active stimulus content, between temporal, spatial, and spatiotemporal visualisations, and between point-based and AOI-based representations of gaze. 

Our approach primarily targets static stimuli, and mostly passive stimulus content. \webveta{} provides a rich set of visualisation views, thus supporting a range of techniques to cover temporal, spatial, and spatiotemporal analysis. Finally, we make heavy use of AOIs to drive the analysis, but still support point-based visualisation as well.

\citet{brehmer2013multi} list 11 categories to characterise task typology. 
\rev{~\citet{kurzhals2017task} present six eye tracking visualization tasks to describe multivariate gaze data analysis and their derived analytical tasks.
To support the elasticity of \webveta{} compared with state-of-the-art,} we aim to support the full range of eye tracking visualisation tasks. 
Using ~\citet{brehmer2013multi}'s typology and~\citet{Blascheck2017}'s taxonomy, \autoref{tab:toolscompare} compares \webveta{} with other visual eye tracking analytics tools---both research and commercial ones.  
It should be noted that the selection of tools for this comparison is based on literature review, and author's collaborations with eye-tracking practitioners actively using the tools. The criteria focus on matrix-based visualisation and interactive group-level exploration. It is not meant to be exhaustive by any means; we have rather chosen a set of tools that cover a range of different functionalities and tasks that is closest to our work. 
\vspace*{1ex}

While existing visualisation techniques come with a variety of visualisation views~\cite{Blascheck:2014:SAV}, we want to pick out matrix-based visualisation as one particularly important one. 
Matrix visualisations can show AOI transitions of fixations and of aggregated or concatenated fixations over multiple samples between AOIs~\cite{wang2021scanpath}, scanpath similarity between samples~\cite{kumar2019clustered,burch2018eyemsa}, relationships of mean fixation duration on AOIs by tasks~\cite{siirtola2009visual}, and so on.

\citet{kurzhals2016eye} point out that only little work had been done in combining traditional numerical-based analysis and human-in-the-loop gaze analytical reasoning. Therefore, there is still need for an improved combination of visual-interactive exploration and advanced visualisations. 

In summary, \webveta{} adopts many of the individual visualisation views already known from the literature. In particular, it makes heavy use of matrix visualisation, aggregation, and filtering. However, \rev{as shown in \autoref{tab:toolscompare},} \webveta{} is unique in its combination of such individual components, the number and extent of supported visualisation views, their integration in the interactive and user-steered process, and the level of exposure of parameters to the user.

\section{System and design}
\label{sec:design}


In the following, we describe key characteristics of our design. The main GUI components are illustrated in~\autoref{fig:webveta_overview}.
Data is loaded from a text file with three tab separated fields: time, gaze x- and y-coordinates, easily obtained from most eye-tracking equipment and therefore generalisable to many eye-tracking applications~\cite{kurzhals2017task}. 
\rev{Additionally, \webveta{} adopts an optional fourth input field where a user can label the timestamp and gaze coordinate with an associated experimental condition which can be mapped to the visual gaze analysis}.




\subsection{Design overview}
\webveta{} consists of five GUI panels. The data management panel in \autoref{fig:webveta_overview}(a) allows \rev{for managing input data across multiple dimensions: \textit{Who} (samples), \textit{Where} (AOIs), and \textit{When} (TWIs)}, described in~\citet{kurzhals2017task}. 
\rev{Furthermore, this panel allows visual aggregation, configuration, and manipulation of grouping across these dimensions. Export of various results of the current sessions such as gaze features, AOI mapping of fixations, and visualisations are available in multiple buttons.} 
Other tabs of the data management panel support text annotation, customising colour schemes using analysts' existing definitions, and export configuration for visual elements such as axis labels or charts. A parameter control panel in \autoref{fig:webveta_overview}(c), including the control bar at the top of \autoref{fig:webveta_overview}(b) allows for toggling and showing current state of data filtering by individual, group level, or overall in each data dimension. This panel also includes an interactive legend for filtering out unselected sample groups.
For example, \rev{\autoref{fig:webveta_overview}(c)--middle shows a legend of 3 sample groups (green, ogange, purple), where a mouse click on green legend resulted in visually filtering out the scanpath and density maps of organe and purple groups in the spatial view, as seen in~\autoref{fig:webveta_overview}(b).}

Three main visualisation panels provide coordinated views for flexible exploration: the metric panel in \autoref{fig:webveta_overview}(d), the spatial panel in \autoref{fig:webveta_overview}(b), and the timeline panel in \autoref{fig:webveta_overview}(e).  The metric panel in \autoref{fig:webveta_overview}(d) provides flexible multivariate data exploration and comparison (\rev{\textit{Relate}~\cite{kurzhals2017task}}).
The metric panel (d) can be switched between a matrix and a histogram view (next to the matrix tab) of fixation and saccade distributions.
A video view, embedded to the metric panel (left of the matrix), as seen in \autoref{fig:webveta_overview}(d), allows a user to attach a video associated with each sample. When a TWI is selected, it navigates the video of a currently selected sample to the corresponding timespan of the TWI for gaining context about eye movement data.
Furthermore, coordinated views showing the spatial panel (b) can be linked with a matrix or histogram view in Panel (d),  which present group or individual attention distribution numerically and across the spatial view field via density maps, with fixations and saccades showing spatial context. 
A scanpath showing temporal eye movement sequence is shown on-demand in~\autoref{fig:webveta_overview}(b) on mouse hovering over the scarfplot of AOI sequences in~\autoref{fig:webveta_overview}(e). 
\vspace{-1em}

\subsection{Multi-way visual exploration with matrix-based overview}
\label{sec:metric_visualisations}
In the metric panel in \autoref{fig:webveta_overview}(d), a matrix shows view metrics between pairs of features grouped by either sample, TWI, or AOI. 
The matrix supports a common range of fixation, saccade and scanpath metrics by \citet{poole2006eye}.
\rev{To support both exploratory and comparative analytical tasks}, the matrix shows similarity comparison (\rev{\textit{Compare}~\cite{kurzhals2017task}}) in terms of visual metrics within each of the data dimension: samples, AOIs, and TWIs. Examples include AOI (group) sequence scores, based on encoding each sample's AOI visiting sequence, and string alignment according to the Needleman-Wunsch algorithm~\cite{needleman1970general}. Similarity of AOI transitions between sample (group) and attention distribution (density overlap of AOIs or density maps).

An example of a controlled experiment dataset using metric panel and spatial panel of 16 samples is shown in~\autoref{fig:teaser}. To take an overview of the entire dataset, an analyst can directly jump into any multivariate data exploration at individual (a,b,c) or group level (b,d,e,f). Temporal filter is applied where metrics are aggregated for all repeated measures within TWI Group 2. To gain spatial context, \autoref{fig:teaser}(f) shows results of brushing and linking over matrix cell (SankeyDiagram1, SankeyDiagram2), denoted as (SD1, SD2) of AOI transition matrix in \autoref{fig:teaser}(c), highlighting AOI transitions in Group 4 has higher frequency of ``looking out'' from SD1 to SD2 than Group 11, supported by fixation context (red, green, yellow dots) where the brushed saccades connect to.

To visually identify gaze patterns (\rev{\textit{Detect}~\cite{kurzhals2017task}}) from visual exploration , visualisation algorithms such as matrix reordering~\cite{behrisch2016matrix}, directional and spatial saccade bundling (in the spatial view)~\cite{hurter2011moleview} are applied to optimise the layout and reveal high-level structures or patterns for better capturing the relationships within the data~\cite{chen2022s}. \rev{\autoref{fig:teaser}(a) shows an example of applying the ``optimal leaf'' ordering to reveal three potential visual blocks (using AOI transition similarity criteria) along the diagonal~\cite{behrisch2016matrix}. 
Samples can then be grouped, while moving into various matrix relationship visualisations to cross-verify the grouping and metrics results}. 




\subsection{Interactive visual grouping and exploration in multiple data dimensions}
\label{sec:interaction_design}
\webveta{} supports a comprehensive set of tools for group-level analysis. 
A user can aggregate metrics of fixations, visitations, density distribution, saccade transition counts, etc, from two or more samples, with spatial aggregation of these metrics in AOI groups, and temporal aggregation of them in TWI groups, identified by unique logical group IDs (GIDs).
\autoref{fig:teaser} shows a temporal aggregation of saccade transition similarity by a selected TWI Group: 2 (defined in~\autoref{sec:case_study}), across samples (\autoref{fig:teaser}(a)), or samples and sample groups (\autoref{fig:teaser}(b)).
In the spatial panel, visual grouping and data granularity also applies to a focus-and-context interaction which allows selecting an AOI and gaining an overview of surrounding spatial context of fixations by samples, spatial, and temporal filtering. \autoref{fig:teaser}(f) shows fixations context: before/after visiting/glancing a selected (focus) AOI (SD1, denoted by bolder boundary). 
If a different TWI group or a different sample filtering is applied, the visuals will show different results; 
This also applies to brushing \& linking where coordinated views highlight selected group's fixation, saccades as one brushes over matrix cells or complementary views. Timeline panel displays AOI group sequence, or scarf plots coloured by AOI groups.
A change of grouping triggers metric re-calculation using data aggregation and the resulting visualisations are re-generated.

\subsection {Implementation}
The software architecture is an MVC framework using HTML canvas for visuals.
Metrics aggregation is achieved via identifying data elements with the same logical group ID (GID).
\webveta{} has native support for having more than one instance, via spawning a new tab within a web browser for a flexible comparison across experimental conditions. 
It allows insights to be disseminated with peers through exporting and restoring the visualisations.

\section{Exploratory and Comparative Analysis Examples}
\label{sec:case_study}
\begin{figure*}
    \centering
    
     \includegraphics[width=0.8\textwidth]{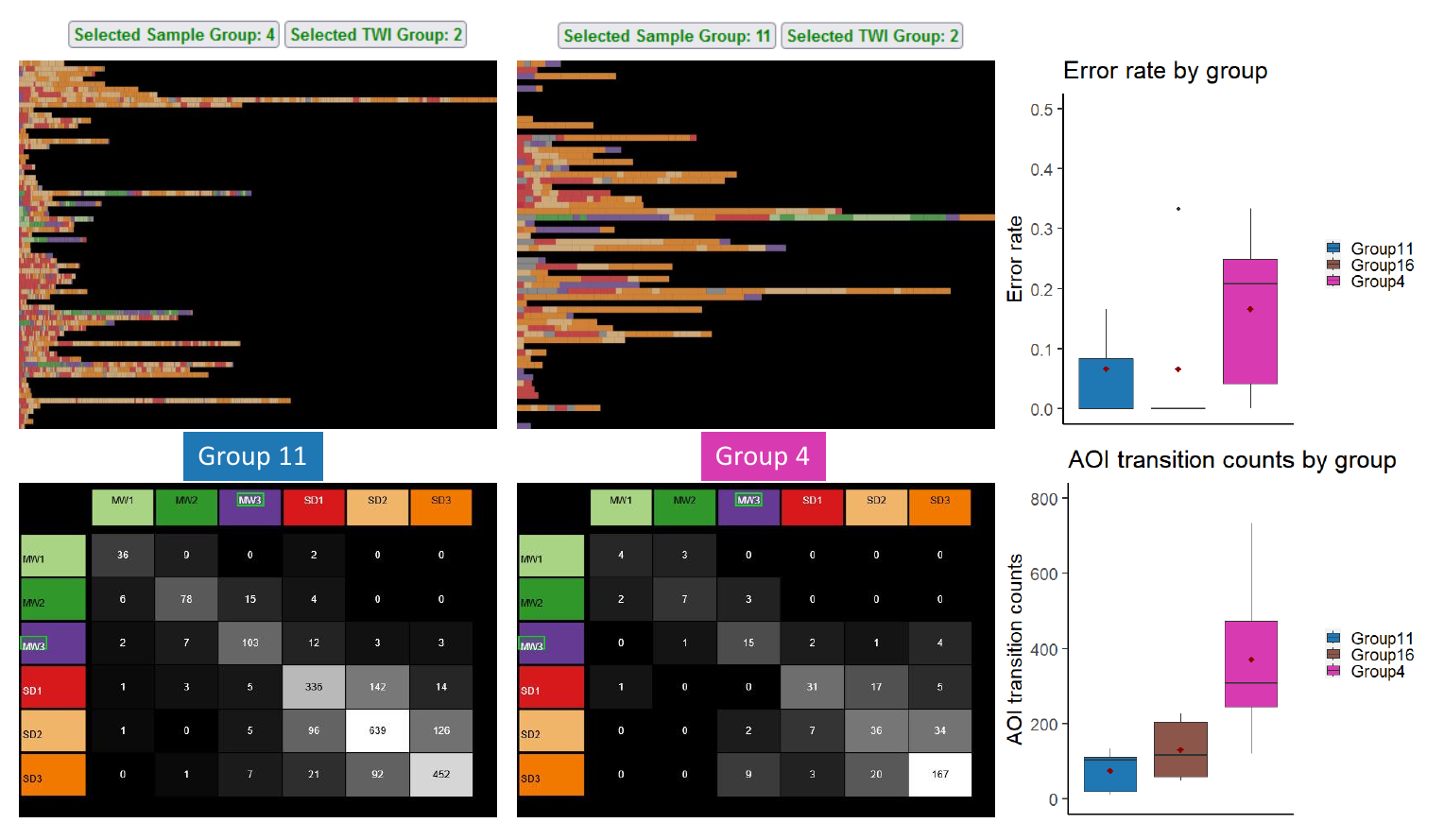}
    
    \Description[]{}
    \caption[]{Visual comparison of AOI transitions between two samples groups: Group 4 and Group 11 identified through multi-way exploration in~\autoref{fig:teaser}. Data is filtered by temporal aggregation: selected TWI Group 2, but it also can be changed to other TWI groups, all TWIs (for gaining an entire overview), or an individual TWI. 
    \label{fig:group_level_matrix_timeline.pdf}
     \vspace{-1em}
    }
\end{figure*}
\begin{figure*}
    \centering
     \includegraphics[width=0.8\textwidth]{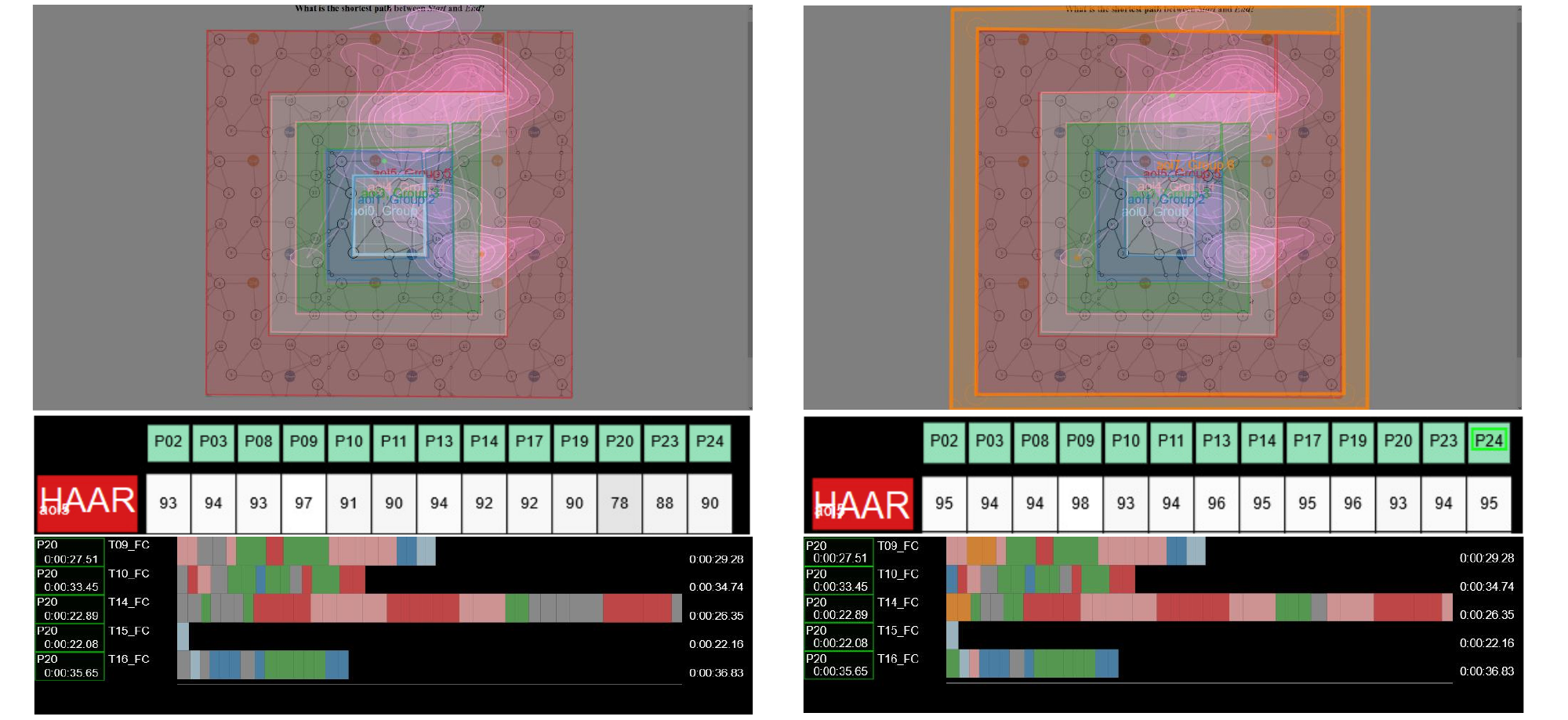}
    
    \Description[]{}
    \caption[]{Example of exploring AOI definitions (using the spatial panel) for labelling fixations, linked with the matrix: sample-AOI relationships with the hit-any-AOI-rate (HAAR) metric, and the timeline panel.
   
    \label{fig:case_study_haar}
     \vspace{-1em}
    }
\end{figure*}
The features of \webveta{} were developed through a series of close collaborations with eye-tracking practitioners actively using the software to analyse their collected study data.  We now present two examples of actual use.  \rev{We refer to the task typology in~\autoref{tab:toolscompare} when describing the activities performed}. 

The first example use case is an evaluation of whether information visualisations placed side-by-side could be complementary~\cite{chang2017evaluating}.
The experimental design involved 24 participants, and 12 repeated measures across 4 tasks (\autoref{fig:teaser}).
\citet{chang2017evaluating} used the Tobii Pro software to label AOIs and extracted fixation metrics, and they built custom visualisations to inspect the results before running statistical comparisons. 
This is a typical approach for the analysis of eye movement data, and here it  succeeded in detecting three eye movement strategies between two large predefined AOIs within the stimuli. However, it was difficult to analyse finer-grained gaze behaviours.


In contrast, with \webveta{}, one can flexibly begin analysis \rev{with any stage of the multi-way exploration (\autoref{sec:metric_visualisations})} from different perspectives. 

\textbf{1. Confirm pre-identified hypothesis}: 
We first \rev{created two AOIs using definitions of the left and right side-by-side diagrams, as described in the prior study (\emph{Import})}~\cite{chang2017evaluating}.
Trials were imported as TWIs representing start time and end time of each trial (\emph{Import}).
By selecting a local filtering with a TWI (selecting a specific trial), we aggregated samples to observe within-subject differences across participants in terms of the fixation percentage time on AOIs metrics \rev{(\emph{Filter}, \emph{Aggregate})}. 
This allowed us to quickly confirm the exclusive and parallel use of the left and right side-by-side views for a representative graph exploration task (i.e., Out-Link task~\cite{chang2017evaluating}).

\textbf{2. Data-driven exploration}: 
To gain further insights into the use of each left and right view, we created new finer-grained AOIs not reported previously by \citet{chang2017evaluating}, by labeling three matrix waves and three Sankey Diagrams within the AOI \rev{(\emph{Annotate})}, as shown in \autoref{fig:teaser}(f). Six AOIs were created and two logical AOI group IDs were assigned to each AOI from the group editor in the data management panel \rev{(\emph{Aggregate}, \emph{Change})}. 



Next, we selected a larger granularity to gain an overview (TWI Group 2), i.e., temporal aggregate of data dimension, which aggregated metrics from all the trials belonging to the Out-Link task \rev{(\emph{Select}, \emph{Navigate})}. 
Using coordinated views of metric, spatial, and timelines, it gave us additional context. In the spatial view in \autoref{fig:webveta_overview}(b), we first inspected the gaze points entering (green dots), leaving (red dots), or glancing out (yellow dots) of the leftmost Sankey Diagram AOI (boundary highlighted in bold when selected from the spatial view or AOI tab). 
Subsequently, we repeated the same exploration by switching between sample groups to allow us to gain insights into each sample group's ``gazing out rate (to the left or right) of the selected leftmost Sankey-diagram (SD1)'' \rev{(\emph{Select}, \emph{Navigate})}. 
As a result, we found that Group 11 has a higher number of red dots than Group 4 (as seen in~\autoref{fig:teaser}(f)).
\textbf{3. Comparative analysis}: 
To compare group behaviours, 
we jumped into data-driven exploration of the stage denoted in \autoref{fig:teaser}(a).
By comparing pairwise AOI transitions similarity and arranging the matrix layout with optimal matrix reordering (\rev{\emph{Arrange}}), three different visual blocks were visually identified along the diagonal. We edited the group ID of each sample according to the block patterns found \rev{(\emph{Aggregate}, \emph{Change})}.
We visually identified that Group 11 and Group 4 had similar AOI group fixation percentage times based on new metrics derived at a group-level (\autoref{fig:teaser}(f, g)) \rev{(\emph{Derive})}. 
For Group 11, \autoref{fig:group_level_matrix_timeline.pdf} showed it has less frequent AOI transitions than Group 4. We extracted AOI transitions metrics of each individual participant from \webveta{} (\rev{\emph{Produce}}), and executed our statistical test script of the extracted data in R . We found a potential weak trend ($p=0.09$) that there was an overall difference in AOI transitions across groups (Friedman's non-parametric test). This provided the new insight that Group 4 participants using Sankey Diagram had a weak trend to have smaller number of AOI transitions than Group 11. During exploration, we also spotted the outlier P21, which has less than 10 fixations. It was then excluded (\rev{\emph{Filter}}) from the rest of the exploration. 

\textbf{4. Coordinated views and interactive exploration}: 
Our second use case is an example of exploratory analysis with data collected from a user study where AOIs are not pre-identified~\cite{chen2020doughnets}.
\webveta{} supports dynamically annotating unlabelled gaze-data with AOIs while visually inspecting the impact of AOI definitions on visual metrics (such as AOI uncertainty metrics), to arrive at more robust AOI definitions. An example of visually improving hit-any-AOI-rate (HAAR) is shown in~\autoref{fig:case_study_haar}, which could lead to higher quality of eye tracking data analysis~\cite{wang2022impact}.
Visual exploration reduces the impact of AOI uncertainty by improving HAAR from 77\% (left) to 88\% (by manipulating the shape of green and pink polygons to be more aligned at the boundaries) \rev{(\emph{Annotate}, \emph{Change})}. Further visual checking and manipulation result in an improved HAAR of 93\% (right) for a more suitable AOI definition. Inspecting the timeline panel (at the bottom right) shows that fixations at the gray area (bottom left) were mapped to AOI4 and AOI6.
\section{conclusion}
We have presented \webveta{}, a new visual eye tracking analytics toolkit that supports \rev{flexible exploratory} gaze analysis.
It combines matrix-based overview with the
ability to aggregate visual metrics at individual or group-level of sample, area-, and time-window of interest, that are dynamically linked with their spatiotemporal visualisations.
\rev{Our comparative table shows that \webveta{} provides a comprehensive and interactive toolkit for exploratory gaze analysis beyond other current toolkits, in terms of a full range of visualisation task taxonomy supported}. 
Eye tracking examples show that with the unified toolkit and multiple coordinated views, it supports multiple types of eye tracking datasets and a wide range of analytical tasks. 


\rev{Furthermore, there has not been many extensive user studies on how a flexible and generic visual eye tracking analytics toolkit affects eye tracking analysts' workflow in the real-world analysis context}. 
With this approach, we hope to further elicit new insights through evaluating with eye tracking experts to propose a more human-centric approach to visual eye tracking analytics.

\begin{acks}
The authors would like to thank Chunlei Chang for actively using the toolkit and provide us his feedback, Ishwari Bhade for co-developing \webveta{}, Kuno Kurzhals for the discussion of visual eye tracking toolkit comparisons, as well as the anonymous reviewers for their helpful comments on this paper.
This research is funded by Monash Faculty of Information Technology and the Deutsche Forschungsgemeinschaft (DFG, German Research Foundation) – Project-ID 251654672 – TRR 161.
\end{acks}


\bibliographystyle{ACM-Reference-Format}

\bibliography{references}


\begin{thebibliography}{30}


\ifx \showCODEN    \undefined \def \showCODEN     #1{\unskip}     \fi
\ifx \showDOI      \undefined \def \showDOI       #1{#1}\fi
\ifx \showISBNx    \undefined \def \showISBNx     #1{\unskip}     \fi
\ifx \showISBNxiii \undefined \def \showISBNxiii  #1{\unskip}     \fi
\ifx \showISSN     \undefined \def \showISSN      #1{\unskip}     \fi
\ifx \showLCCN     \undefined \def \showLCCN      #1{\unskip}     \fi
\ifx \shownote     \undefined \def \shownote      #1{#1}          \fi
\ifx \showarticletitle \undefined \def \showarticletitle #1{#1}   \fi
\ifx \showURL      \undefined \def \showURL       {\relax}        \fi
\providecommand\bibfield[2]{#2}
\providecommand\bibinfo[2]{#2}
\providecommand\natexlab[1]{#1}
\providecommand\showeprint[2][]{arXiv:#2}

\bibitem[Afzal et~al\mbox{.}(2022)]%
        {afzal2022investigating}
\bibfield{author}{\bibinfo{person}{Umair Afzal}, \bibinfo{person}{Arnaud
  Prouzeau}, \bibinfo{person}{Lee Lawrence}, \bibinfo{person}{Tim Dwyer},
  \bibinfo{person}{Saikiranrao Bichinepally}, \bibinfo{person}{Ariel Liebman},
  {and} \bibinfo{person}{Sarah Goodwin}.} \bibinfo{year}{2022}\natexlab{}.
\newblock \showarticletitle{Investigating Cognitive Load in Energy Network
  Control Rooms: Recommendations for Future Designs}.
\newblock \bibinfo{journal}{\emph{Frontiers in Psychology}}
  \bibinfo{volume}{13} (\bibinfo{year}{2022}).
\newblock


\bibitem[Behrisch et~al\mbox{.}(2016)]%
        {behrisch2016matrix}
\bibfield{author}{\bibinfo{person}{Michael Behrisch}, \bibinfo{person}{Benjamin
  Bach}, \bibinfo{person}{Nathalie Henry~Riche}, \bibinfo{person}{Tobias
  Schreck}, {and} \bibinfo{person}{Jean-Daniel Fekete}.}
  \bibinfo{year}{2016}\natexlab{}.
\newblock \showarticletitle{Matrix Reordering Methods for Table and Network
  Visualization}.
\newblock \bibinfo{journal}{\emph{Computer Graphics Forum}}
  \bibinfo{volume}{35}, \bibinfo{number}{3} (\bibinfo{year}{2016}),
  \bibinfo{pages}{693--716}.
\newblock


\bibitem[{Blascheck} et~al\mbox{.}(2016)]%
        {Blascheck2016}
\bibfield{author}{\bibinfo{person}{T. {Blascheck}}, \bibinfo{person}{M.
  {John}}, \bibinfo{person}{K. {Kurzhals}}, \bibinfo{person}{S. {Koch}}, {and}
  \bibinfo{person}{T. {Ertl}}.} \bibinfo{year}{2016}\natexlab{}.
\newblock \showarticletitle{{VA$^2$}: A Visual Analytics Approach for
  Evaluating Visual Analytics Applications}.
\newblock \bibinfo{journal}{\emph{IEEE Transactions on Visualization and
  Computer Graphics}} \bibinfo{volume}{22}, \bibinfo{number}{1}
  (\bibinfo{year}{2016}), \bibinfo{pages}{61--70}.
\newblock


\bibitem[Blascheck et~al\mbox{.}(2014)]%
        {Blascheck:2014:SAV}
\bibfield{author}{\bibinfo{person}{T. Blascheck}, \bibinfo{person}{K.
  Kurzhals}, \bibinfo{person}{M. Raschke}, \bibinfo{person}{M. Burch},
  \bibinfo{person}{D. Weiskopf}, {and} \bibinfo{person}{T. Ertl}.}
  \bibinfo{year}{2014}\natexlab{}.
\newblock \showarticletitle{{State-of-the-Art of Visualization for Eye Tracking
  Data}}. In \bibinfo{booktitle}{\emph{EuroVis -- STARs}},
  \bibfield{editor}{\bibinfo{person}{R.~Borgo},
  \bibinfo{person}{R.~Maciejewski}, {and} \bibinfo{person}{I.~Viola}} (Eds.).
  \bibinfo{publisher}{The Eurographics Association}, \bibinfo{pages}{63--82}.
\newblock


\bibitem[Blascheck et~al\mbox{.}(2017)]%
        {Blascheck2017}
\bibfield{author}{\bibinfo{person}{T. Blascheck}, \bibinfo{person}{K.
  Kurzhals}, \bibinfo{person}{M. Raschke}, \bibinfo{person}{M. Burch},
  \bibinfo{person}{D. Weiskopf}, {and} \bibinfo{person}{T. Ertl}.}
  \bibinfo{year}{2017}\natexlab{}.
\newblock \showarticletitle{Visualization of Eye Tracking Data: A Taxonomy and
  Survey}.
\newblock \bibinfo{journal}{\emph{Computer Graphics Forum}}
  \bibinfo{volume}{36}, \bibinfo{number}{8} (\bibinfo{year}{2017}),
  \bibinfo{pages}{260--284}.
\newblock


\bibitem[Brehmer and Munzner(2013)]%
        {brehmer2013multi}
\bibfield{author}{\bibinfo{person}{Matthew Brehmer} {and}
  \bibinfo{person}{Tamara Munzner}.} \bibinfo{year}{2013}\natexlab{}.
\newblock \showarticletitle{A Multi-Level Typology of Abstract Visualization
  Tasks}.
\newblock \bibinfo{journal}{\emph{IEEE Transactions on Visualization and
  Computer Graphics}} \bibinfo{volume}{19}, \bibinfo{number}{12}
  (\bibinfo{year}{2013}), \bibinfo{pages}{2376--2385}.
\newblock


\bibitem[Burch(2022)]%
        {burch2022eye}
\bibfield{author}{\bibinfo{person}{Michael Burch}.}
  \bibinfo{year}{2022}\natexlab{}.
\newblock \bibinfo{booktitle}{\emph{Eye Tracking and Visual Analytics}}.
\newblock \bibinfo{publisher}{CRC Press}.
\newblock


\bibitem[Burch et~al\mbox{.}(2019)]%
        {burch2019interactive}
\bibfield{author}{\bibinfo{person}{Michael Burch}, \bibinfo{person}{Ayush
  Kumar}, {and} \bibinfo{person}{Neil Timmermans}.}
  \bibinfo{year}{2019}\natexlab{}.
\newblock \showarticletitle{An Interactive Web-Based Visual Analytics Tool for
  Detecting Strategic Eye Movement Patterns}. In
  \bibinfo{booktitle}{\emph{Proceedings of the 11th ACM Symposium on Eye
  Tracking Research \& Applications}}. Article \bibinfo{articleno}{93},
  \bibinfo{numpages}{5}~pages.
\newblock


\bibitem[Burch et~al\mbox{.}(2018)]%
        {burch2018eyemsa}
\bibfield{author}{\bibinfo{person}{Michael Burch}, \bibinfo{person}{Kuno
  Kurzhals}, \bibinfo{person}{Niklas Kleinhans}, {and} \bibinfo{person}{Daniel
  Weiskopf}.} \bibinfo{year}{2018}\natexlab{}.
\newblock \showarticletitle{{EyeMSA:} Exploring Eye Movement Data with Pairwise
  and Multiple Sequence Alignment}. In \bibinfo{booktitle}{\emph{Proceedings of
  the 2018 ACM Symposium on Eye Tracking Research \& Applications}}. Article
  \bibinfo{articleno}{52}, \bibinfo{numpages}{5}~pages.
\newblock


\bibitem[Cai et~al\mbox{.}(2022)]%
        {cai2022towards}
\bibfield{author}{\bibinfo{person}{Minghao Cai}, \bibinfo{person}{Bin Zheng},
  {and} \bibinfo{person}{Carrie Demmans~Epp}.} \bibinfo{year}{2022}\natexlab{}.
\newblock \showarticletitle{Towards Supporting Adaptive Training of Injection
  Procedures: Detecting Differences in the Visual Attention of Nursing Students
  and Experts}. In \bibinfo{booktitle}{\emph{Proceedings of the 30th ACM
  Conference on User Modeling, Adaptation and Personalization}}.
  \bibinfo{pages}{286--294}.
\newblock


\bibitem[Chang et~al\mbox{.}(2017)]%
        {chang2017evaluating}
\bibfield{author}{\bibinfo{person}{Chunlei Chang}, \bibinfo{person}{Benjamin
  Bach}, \bibinfo{person}{Tim Dwyer}, {and} \bibinfo{person}{Kim Marriott}.}
  \bibinfo{year}{2017}\natexlab{}.
\newblock \showarticletitle{Evaluating Perceptually Complementary Views for
  Network Exploration Tasks}. In \bibinfo{booktitle}{\emph{Proceedings of the
  2017 CHI Conference on Human Factors in Computing Systems}}.
  \bibinfo{pages}{1397--1407}.
\newblock


\bibitem[Chang et~al\mbox{.}(2018)]%
        {chang2018evaluation}
\bibfield{author}{\bibinfo{person}{Chunlei Chang}, \bibinfo{person}{Tim Dwyer},
  {and} \bibinfo{person}{Kim Marriott}.} \bibinfo{year}{2018}\natexlab{}.
\newblock \showarticletitle{An Evaluation of Perceptually Complementary Views
  for Multivariate Data}. In \bibinfo{booktitle}{\emph{Proceedings of the 2018
  IEEE Pacific Visualization Symposium (PacificVis)}}. IEEE,
  \bibinfo{pages}{195--204}.
\newblock


\bibitem[Chen(2022)]%
        {chen2022s}
\bibfield{author}{\bibinfo{person}{Kun-Ting Chen}.}
  \bibinfo{year}{2022}\natexlab{}.
\newblock \showarticletitle{It's a Wrap! Visualisations that Wrap Around
  Cylindrical, Toroidal, or Spherical Topologies}.
\newblock \bibinfo{journal}{\emph{arXiv preprint arXiv:2209.13251}}
  (\bibinfo{year}{2022}).
\newblock


\bibitem[Chen et~al\mbox{.}(2020)]%
        {chen2020doughnets}
\bibfield{author}{\bibinfo{person}{Kun-Ting Chen}, \bibinfo{person}{Tim Dwyer},
  \bibinfo{person}{Kim Marriott}, {and} \bibinfo{person}{Benjamin Bach}.}
  \bibinfo{year}{2020}\natexlab{}.
\newblock \showarticletitle{DoughNets: Visualising Networks Using Torus
  Wrapping}. In \bibinfo{booktitle}{\emph{Proceedings of the 2020 CHI
  Conference on Human Factors in Computing Systems}}. Article
  \bibinfo{articleno}{53}, \bibinfo{numpages}{11}~pages.
\newblock


\bibitem[Chen et~al\mbox{.}(2023)]%
        {chen2023reading}
\bibfield{author}{\bibinfo{person}{Kun-Ting Chen}, \bibinfo{person}{Quynh~Quang
  Ngo}, \bibinfo{person}{Kuno Kurzhals}, \bibinfo{person}{Kim Marriott},
  \bibinfo{person}{Tim Dwyer}, \bibinfo{person}{Michael Sedlmair}, {and}
  \bibinfo{person}{Daniel Weiskopf}.} \bibinfo{year}{2023}\natexlab{}.
\newblock \bibinfo{title}{Reading Strategies for Graph Visualizations that Wrap
  Around in Torus Topology}.
\newblock
\newblock
\showeprint[arxiv]{2303.17066}~[cs.HC]


\bibitem[Hurter et~al\mbox{.}(2011)]%
        {hurter2011moleview}
\bibfield{author}{\bibinfo{person}{Christophe Hurter},
  \bibinfo{person}{Alexandru Telea}, {and} \bibinfo{person}{Ozan Ersoy}.}
  \bibinfo{year}{2011}\natexlab{}.
\newblock \showarticletitle{MoleView: An Attribute and Structure-Based Semantic
  Lens for Large Element-Based Plots}.
\newblock \bibinfo{journal}{\emph{IEEE Transactions on Visualization and
  Computer Graphics}} \bibinfo{volume}{17}, \bibinfo{number}{12}
  (\bibinfo{year}{2011}), \bibinfo{pages}{2600--2609}.
\newblock


\bibitem[Kumar et~al\mbox{.}(2019)]%
        {kumar2019clustered}
\bibfield{author}{\bibinfo{person}{Ayush Kumar}, \bibinfo{person}{Neil
  Timmermans}, \bibinfo{person}{Michael Burch}, {and} \bibinfo{person}{Klaus
  Mueller}.} \bibinfo{year}{2019}\natexlab{}.
\newblock \showarticletitle{Clustered Eye Movement Similarity Matrices}. In
  \bibinfo{booktitle}{\emph{Proceedings of the 11th ACM Symposium on Eye
  Tracking Research \& Applications}}. Article \bibinfo{articleno}{82},
  \bibinfo{numpages}{9}~pages.
\newblock


\bibitem[Kurzhals et~al\mbox{.}(2017)]%
        {kurzhals2017task}
\bibfield{author}{\bibinfo{person}{Kuno Kurzhals}, \bibinfo{person}{Michael
  Burch}, \bibinfo{person}{Tanja Blascheck}, \bibinfo{person}{Gennady
  Andrienko}, \bibinfo{person}{Natalia Andrienko}, {and}
  \bibinfo{person}{Daniel Weiskopf}.} \bibinfo{year}{2017}\natexlab{}.
\newblock \showarticletitle{A Task-Based View on the Visual Analysis of
  Eye-Tracking Data}.
\newblock In \bibinfo{booktitle}{\emph{Eye Tracking and Visualization}},
  \bibfield{editor}{\bibinfo{person}{Michael Burch}, \bibinfo{person}{Lewis
  Chuang}, \bibinfo{person}{Brian Fisher}, \bibinfo{person}{Albrecht Schmidt},
  {and} \bibinfo{person}{Daniel Weiskopf}} (Eds.). \bibinfo{publisher}{Springer
  International Publishing}, \bibinfo{address}{Cham}, \bibinfo{pages}{3--22}.
\newblock


\bibitem[Kurzhals et~al\mbox{.}(2016a)]%
        {kurzhals2016eye}
\bibfield{author}{\bibinfo{person}{Kuno Kurzhals}, \bibinfo{person}{Brian
  Fisher}, \bibinfo{person}{Michael Burch}, {and} \bibinfo{person}{Daniel
  Weiskopf}.} \bibinfo{year}{2016}\natexlab{a}.
\newblock \showarticletitle{Eye tracking evaluation of visual analytics}.
\newblock \bibinfo{journal}{\emph{Information Visualization}}
  \bibinfo{volume}{15}, \bibinfo{number}{4} (\bibinfo{year}{2016}),
  \bibinfo{pages}{340--358}.
\newblock


\bibitem[Kurzhals et~al\mbox{.}(2016b)]%
        {kurzhals2016visual}
\bibfield{author}{\bibinfo{person}{Kuno Kurzhals}, \bibinfo{person}{Marcel
  Hlawatsch}, \bibinfo{person}{Christof Seeger}, {and} \bibinfo{person}{Daniel
  Weiskopf}.} \bibinfo{year}{2016}\natexlab{b}.
\newblock \showarticletitle{Visual Analytics for Mobile Eye Tracking}.
\newblock \bibinfo{journal}{\emph{IEEE Transactions on Visualization and
  Computer Graphics}} \bibinfo{volume}{23}, \bibinfo{number}{1}
  (\bibinfo{year}{2016}), \bibinfo{pages}{301--310}.
\newblock


\bibitem[Kwok et~al\mbox{.}(2019)]%
        {kwok2019gaze}
\bibfield{author}{\bibinfo{person}{Tiffany C.~K. Kwok}, \bibinfo{person}{Peter
  Kiefer}, \bibinfo{person}{Victor~R. Schinazi}, \bibinfo{person}{Benjamin
  Adams}, {and} \bibinfo{person}{Martin Raubal}.}
  \bibinfo{year}{2019}\natexlab{}.
\newblock \showarticletitle{Gaze-Guided Narratives: Adapting Audio Guide
  Content to Gaze in Virtual and Real Environments}. In
  \bibinfo{booktitle}{\emph{Proceedings of the 2019 CHI Conference on Human
  Factors in Computing Systems}}. Article \bibinfo{articleno}{491},
  \bibinfo{numpages}{12}~pages.
\newblock


\bibitem[Menges et~al\mbox{.}(2020)]%
        {menges2020visualization}
\bibfield{author}{\bibinfo{person}{Raphael Menges}, \bibinfo{person}{Sophia
  Kramer}, \bibinfo{person}{Stefan Hill}, \bibinfo{person}{Marius
  Nisslmueller}, \bibinfo{person}{Chandan Kumar}, {and}
  \bibinfo{person}{Steffen Staab}.} \bibinfo{year}{2020}\natexlab{}.
\newblock \showarticletitle{A Visualization Tool for Eye Tracking Data Analysis
  in the Web}. In \bibinfo{booktitle}{\emph{Proceedings of the ACM Symposium on
  Eye Tracking Research and Applications (ETRA '20 Short Papers)}}. Article
  \bibinfo{articleno}{46}, \bibinfo{numpages}{5}~pages.
\newblock


\bibitem[Needleman and Wunsch(1970)]%
        {needleman1970general}
\bibfield{author}{\bibinfo{person}{Saul~B. Needleman} {and}
  \bibinfo{person}{Christian~D. Wunsch}.} \bibinfo{year}{1970}\natexlab{}.
\newblock \showarticletitle{A General Method Applicable to the Search for
  Similarities in the Amino Acid Sequence of Two Proteins}.
\newblock \bibinfo{journal}{\emph{Journal of Molecular Biology}}
  \bibinfo{volume}{48}, \bibinfo{number}{3} (\bibinfo{year}{1970}),
  \bibinfo{pages}{443--453}.
\newblock


\bibitem[Panetta et~al\mbox{.}(2020)]%
        {panetta2020iseecolor}
\bibfield{author}{\bibinfo{person}{Karen Panetta}, \bibinfo{person}{Qianwen
  Wan}, \bibinfo{person}{Srijith Rajeev}, \bibinfo{person}{Aleksandra
  Kaszowska}, \bibinfo{person}{Aaron~L Gardony}, \bibinfo{person}{Kevin
  Naranjo}, \bibinfo{person}{Holly~A Taylor}, {and} \bibinfo{person}{Sos
  Agaian}.} \bibinfo{year}{2020}\natexlab{}.
\newblock \showarticletitle{ISeeColor: Method for Advanced Visual Analytics of
  Eye Tracking Data}.
\newblock \bibinfo{journal}{\emph{IEEE Access}}  \bibinfo{volume}{8}
  (\bibinfo{year}{2020}), \bibinfo{pages}{52278--52287}.
\newblock


\bibitem[Poole and Ball(2006)]%
        {poole2006eye}
\bibfield{author}{\bibinfo{person}{Alex Poole} {and} \bibinfo{person}{Linden~J.
  Ball}.} \bibinfo{year}{2006}\natexlab{}.
\newblock \showarticletitle{Eye Tracking in HCI and Usability Research}.
\newblock In \bibinfo{booktitle}{\emph{Encyclopedia of Human Computer
  Interaction}}. \bibinfo{publisher}{IGI global}, \bibinfo{pages}{211--219}.
\newblock


\bibitem[Pozdniakov et~al\mbox{.}(2023)]%
        {pozdniakov2023teachers}
\bibfield{author}{\bibinfo{person}{Stanislav Pozdniakov},
  \bibinfo{person}{Roberto Martinez-Maldonado}, \bibinfo{person}{Yi-Shan Tsai},
  \bibinfo{person}{Vanessa Echeverria}, \bibinfo{person}{Namrata Srivastava},
  {and} \bibinfo{person}{Dragan Gasevic}.} \bibinfo{year}{2023}\natexlab{}.
\newblock \showarticletitle{How Do Teachers Use Dashboards Enhanced with Data
  Storytelling Elements According to their Data Visualisation Literacy
  Skills?}. In \bibinfo{booktitle}{\emph{LAK23: 13th International Learning
  Analytics and Knowledge Conference}}. \bibinfo{pages}{89--99}.
\newblock


\bibitem[Servais et~al\mbox{.}(2022)]%
        {servais2022attentional}
\bibfield{author}{\bibinfo{person}{Ana{\"i}s Servais},
  \bibinfo{person}{Christophe Hurter}, {and} \bibinfo{person}{Emmanuel~J.
  Barbeau}.} \bibinfo{year}{2022}\natexlab{}.
\newblock \bibinfo{title}{Attentional switch to memory: an early and critical
  phase of the cognitive cascade allowing autobiographical memory retrieval}.
\newblock
\newblock
\urldef\tempurl%
\url{https://doi.org/10.31234/osf.io/z32qe}
\showDOI{\tempurl}
\newblock
\shownote{PsyArXiv}.


\bibitem[Siirtola et~al\mbox{.}(2009)]%
        {siirtola2009visual}
\bibfield{author}{\bibinfo{person}{Harri Siirtola}, \bibinfo{person}{Tuuli
  Laivo}, \bibinfo{person}{Tomi Heimonen}, {and} \bibinfo{person}{Kari-Jouko
  R{\"a}ih{\"a}}.} \bibinfo{year}{2009}\natexlab{}.
\newblock \showarticletitle{Visual Perception of Parallel Coordinate
  Visualizations}. In \bibinfo{booktitle}{\emph{Proceedings of the 2009 13th
  International Conference Information Visualisation}}. IEEE,
  \bibinfo{pages}{3--9}.
\newblock


\bibitem[Wang et~al\mbox{.}(2021)]%
        {wang2021scanpath}
\bibfield{author}{\bibinfo{person}{Yao Wang}, \bibinfo{person}{Mihai B{\^a}ce},
  {and} \bibinfo{person}{Andreas Bulling}.} \bibinfo{year}{2021}\natexlab{}.
\newblock \showarticletitle{Scanpath Prediction on Information Visualisations}.
\newblock \bibinfo{journal}{\emph{arXiv preprint arXiv:2112.02340}}
  (\bibinfo{year}{2021}).
\newblock


\bibitem[Wang et~al\mbox{.}(2022)]%
        {wang2022impact}
\bibfield{author}{\bibinfo{person}{Yao Wang}, \bibinfo{person}{Maurice Koch},
  \bibinfo{person}{Mihai B{\^a}ce}, \bibinfo{person}{Daniel Weiskopf}, {and}
  \bibinfo{person}{Andreas Bulling}.} \bibinfo{year}{2022}\natexlab{}.
\newblock \showarticletitle{Impact of Gaze Uncertainty on AOIs in Information
  Visualisations}. In \bibinfo{booktitle}{\emph{Proceedings of the 2022
  Symposium on Eye Tracking Research and Applications}}. Article
  \bibinfo{articleno}{60}, \bibinfo{numpages}{6}~pages.
\newblock


\end{thebibliography}

\end{document}